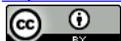
Scientific
Research
Publishing

# Reduction of Anisotropic Volume Expansion and the Optimization of Specific Charge Capacity in Lithiated Silicon Nanowires

## Donald C. Boone

Nanoscience Research Institute, College Park, Maryland, USA
Email: db2585@caa.columbia.edu

## Abstract

This computational research study analyzes the increase of the specific charge capacity that comes with the reduction of the anisotropic volume expansion during lithium ion insertion within silicon nanowires. This research paper is a continuation from previous work that studied the expansion rate and volume increase. It has been determined that when the lithium ion concentration is decreased by regulating the amount of Li ion flux, the lithium ions to silicon atoms ratio, represented by $x$, decreases within the amorphous lithiated silicon (a-$Li_xSi$) material. This results in a decrease in the volumetric strain of the lithiated silicon nanowire as well as a reduction in Maxwell stress that was calculated and Young's elastic module that was measured experimentally using nanoindentation. The conclusion as will be seen is that as there is a decrease in lithium ion concentration there is a corresponding decrease in anisotropic volume and a resulting increase in specific charge capacity. In fact the amplification of the electromagnetic field due to the electron flux that created detrimental effects for a fully lithiated silicon nanowire at $x = 3.75$ which resulted in over a 300% volume expansion becomes beneficial with the decrease in lithium ion flux as $x$ approaches 0.75, which leads to a marginal volume increase of ~25 percent. This could lead to the use of crystalline silicon, c-Si, as an anode material that has been demonstrated in many previous research works to be ten times greater charge capacity than carbon base anode material for lithium ion batteries.

## 1. Introduction

The lithiated silicon nanowire has the potential of being a great advancement in anode material in lithium ion batteries (LIBs) [1]. As well documented in the current literature of research on LIBs, the specific charge capacity (scc) of lithiated silicon has been measured to be greater than 10 times that of carbon base anode batteries [2]. Unfortunately the overwhelming volume expansion in excess of 300% of the silicon nanowire during lithiation has made this material ineffective for the future of lithium ion batteries due to the resulting fracture and failure of this material. This volume expansion appears to occur during full lithiation described by the lithium-silicon material $Li_xSi$ where $x = 3.75$ defines the state where the liathiated silicon nanowire is at full lithiation. When this occurs the volume expands at an uncontrollable anisotropic rate where the $\langle 110 \rangle$ crystallography direction could possibly increase 12 times greater than that of the $\langle 111 \rangle$ orthogonal direction within the silicon nanowire. In this computation research work it will be demonstrated that this volume expansion could be avoided if the lithium ion flux rate is decreasing where $x < 3.75$. The paper will draw heavily on the previous work done by Boone [3] [4], however information will come from other sources that will be indispensable in the calculation of volumetric strain and hence the volume expansion.

The research body of knowledge for lithiated silicon anode materials has been exclusively focused on lithium ion diffusion process. The research work by Liu *et al.* anisotropic volume expansion of lithiated silicon nanowires was studied employing transmission electron microscope (TEM) and electron diffraction pattern (EDP) [1] [5]. In this study, a morphology evolution of the lithiated silicon nanowire started from a pristine crystalline silicon (c-Si) nanowire at 155 nm in diameter prior to lithium insertion to a 17% diameter increase in the $\langle 111 \rangle$ direction and a 170% increase in diameter of 485 nm was measured in the $\langle 110 \rangle$ direction after full lithiation. In addition, a crack along the longitudinal direction of $\langle 112 \rangle$ was detected. A similar result was performed by Yang *et al.* by utilizing a chemomechanical finite element model to simulate several crystallography orientation-dependent anisotropic volume expansions of over 300%, with increases initiating at the interfacial reaction fronts of lithiated silicon nanowire models [6]. The research study performed by Cubuk *et al.* used the kinetic Monte Carlo (kMC) method to simulate the lithium atoms insertion into silicon nanowire. The expansion rates were calculated seven times faster in the $\langle 110 \rangle$ direction than the $\langle 111 \rangle$ direction [7]. This gave the expanded lithiated silicon nanowire a described "dumbbell cross section" shape that resembled the Cassini oval curve geometry [8]. The work drawn from Jung *et al.* molecular dynamics/density functional theory (MD/DFT) simulation was created to study the atomistic behavior of the two-phase interfacial reaction front barrier that separates the c-Si and $Li_xSi$ material [9]. All of these works support the findings that the volume of silicon nanowires during lithium atom/ion insertion will increase non-isotopically by nature.

For the purpose of continuity, our theoretical apparatus will be briefly re-examined as it was presented in our previous work. Prior to the beginning of the lithiation process, the individual lithium atoms are ionized reducing them to the constitutive particles of lithium ions and free electrons [10]. A constant voltage of 2 V is applied to an electric series circuit in order for the lithiation process to begin. The electrons and lithium ions will enter the silicon nanowire at opposing ends and therefore travel in opposite directions (**Figure 1(a)**). When the lithiation begins, this initiates a process of transforming the silicon from c-Si to an amorphous lithiated silicon (a-LiSi) matrix [11]. The free electrons or electron flux varies and increases with the continue diffusion of lithium ions within the silicon nanowire [12]. Since the electrons are moving charge particles, they are the source of the applied quantized electromagnetic field. The geometric model that will be the basis of our mathematical framework is a diamond crystal silicon lattice, which is composed of eight silicon atoms (**Figure 1(b)**). The silicon lattice is fully lithiated with 30 lithium ions [13].

## 2. Analysis

For this research work, there will be a special notation that will be used throughout this study to indicate the orthogonal directions that are essential element in the presentation of this paper. As an example, the mathematical variables and functions that have directional characteristics will have subscripted notations that will indicate which orthogonal direction is being represented. For an example:

$$A_{ij} = A_{\langle \text{Orthogonal direction} \rangle} \quad i = j = 1 \ \text{ or } \ 2 \ \text{ or } \ 3; \tag{1}$$

$$A_{11} = A_{\langle 110 \rangle}$$

$$A_{22} = A_{\langle 111 \rangle}$$

$$A_{33} = A_{\langle 112 \rangle}$$

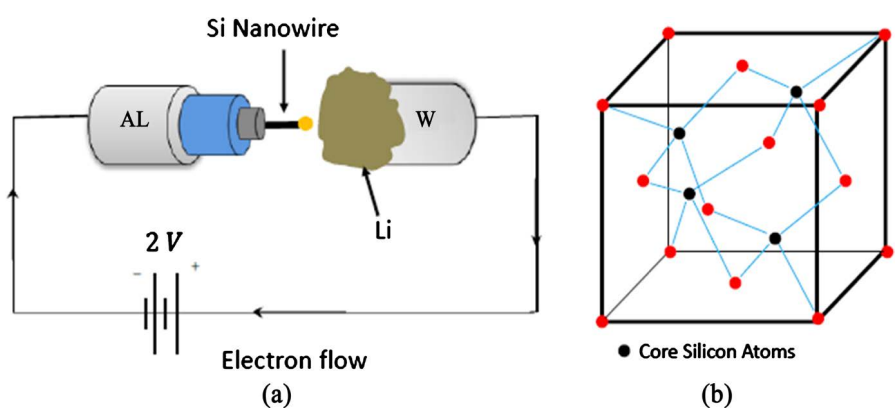

**Figure 1.** (a) *In situ* experimental arrangement for a solid electrochemical cell using lithium metal counter electrode (W); (b) Silicon is a diamond crystalline cubic structure made up of tetrahedral molecules with its hybridized $sp^3$ orbitals within their valence shells filled with covalent bonding electrons from neighboring silicon atoms. Reprinted from AIP Advances 6, 125,027 (2016) American Institute of Physics. Copyright by the author under the terms and conditions of the Creative Commons Attribution (CC BY 4.0).

There will be two main independent variables that will be utilized throughout this work. As previously discuss the independent variable $x$ in $Li_xSi$ will be the lithium ion concentration for lithiated silicon where $x$ = the ratio between lithium ion and silicon atoms. The second independent variable will be known as the average negative charge differential $\bar{n}_c$ define as the difference between the numbers of negative free electrons or electron flux and the number of positive lithium ions or lithium ion flux per unit volume that flows into our computational model that is being studied.

The volume expansion and geometric configuration of the silicon nanowire at full lithiation can best be demonstrated by deriving a set of equations from the Cassini oval geometry [8]. In this study, the computational model simulated the nanowire volume increase slightly above 300% upon the conclusion of lithium ion insertion. There was a volume change $\Delta V_{ij}$ in each of the three orthogonal directions of $\langle 110 \rangle$, $\langle 111 \rangle$ and $\langle 112 \rangle$. However, this study will focus exclusively on $\Delta V_{11}$ since it was calculated that approximately 96% of the volume increase was in that $\langle 110 \rangle$ direction.

$$\Delta V_{\langle 110 \rangle} = \Delta V_{11} = V_{max} \frac{\Delta r_{11}^2}{\Delta r_{max}^2} \tag{2}$$

with $V_{max}$ being the total maximum volume increase and is define as
$V_{max} = \Delta V_{11} + \Delta V_{22} + \Delta V_{33}$ (Figure 2). In terms of the maximum increase in volume, $V_{max}$ can be thought of as being approximately equivalent to $\Delta V_{11}$ at $\bar{n}_c = 6$ because there is a negligible volume expansion in the $\langle 111 \rangle$ and $\langle 112 \rangle$ directions. The three volumetric strains are defined $\varepsilon_{11}, \varepsilon_{22}, \varepsilon_{33}$ in each of the orthogonal directions. The definition of $\Delta r_{11}$ is the decrease in the length of the transition state vector $r_{ij}$ that can be further explained in [4].

$$V_{max} = \varepsilon_{11} + \varepsilon_{22} + \varepsilon_{33} \tag{3}$$

$$V_{max} \approx \Delta V_{11} = \varepsilon_{11} \quad \varepsilon_{22} \approx \varepsilon_{33} \approx 0 \quad \text{at} \quad \bar{n}_c = 6 \tag{4}$$

Since there is only one primary direction that will be analyzed, namely $\langle 110 \rangle$ direction, therefore Young's Modulus $\mathbb{Y}_{Li_xSi}$ will be used in the calculation of volumetric strain $\varepsilon_{11}$.

$$\varepsilon_{11} = \frac{\mathbb{E}_{11}}{\mathbb{Y}_{Li_xSi}} \tag{5}$$

where $\mathbb{E}_{11}$ is the Maxwell stress equation defined in [3] [4]. The determination of Young's Modulus $\mathbb{Y}_{Li_xSi}$ came from experimental data in the research paper Wang *et al.* [14] where a nanoindentation apparatus was used to measure $\mathbb{Y}_{Li_xSi}$ at varies concentration $x$ values under dry and wet conditions. The values used in this research paper under the lithiated dry conditions are as follows:

## 3. Results

The displays in Figure 3 are the volume changes at different lithium ion concentration levels defined by $x$ [15]. As $x$ decreases so does the maximum volume at $\bar{n}_c = 6$. The definition of the current within the lithiated silicon nanowire is

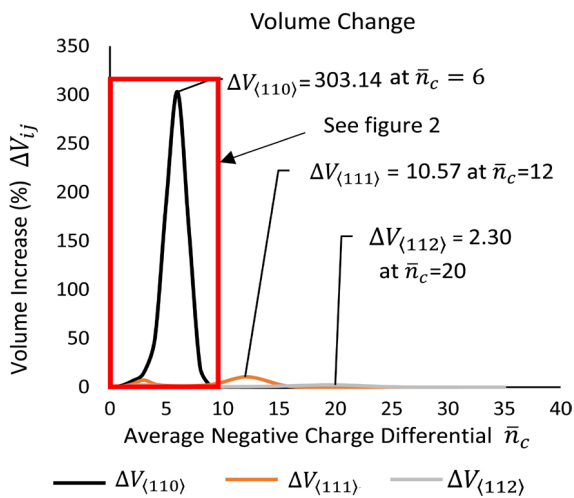

**Figure 2.** Most of the volume increase is in the $\langle 110 \rangle$ direction as displayed in the red rectangle area. Reprinted from Mathematical and Computational Applications—MDPI, October 2017. Copyright by the author under the terms and conditions of the Creative Commons Attribution (CC BY 4.0).

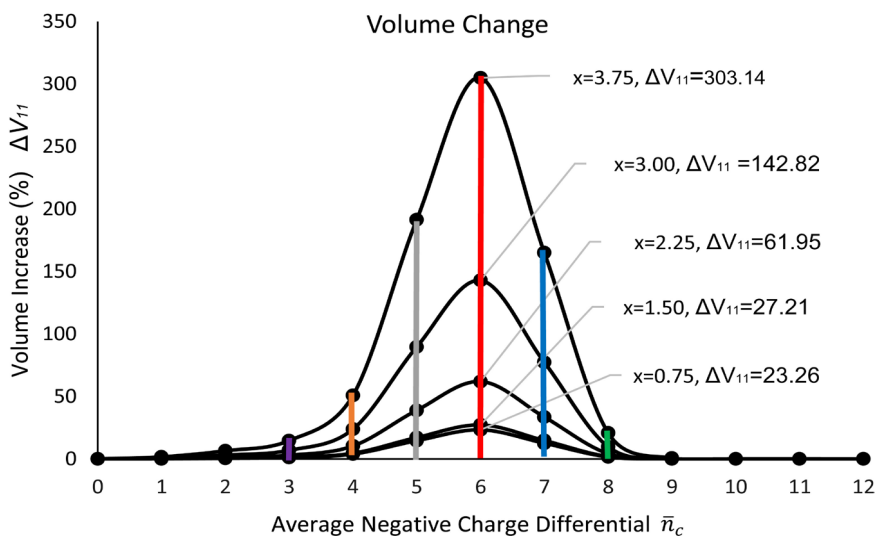

| Li$_x$Si concentration | $x = 0.75$ | $x = 1.00$ | $x = 1.50$ | $x = 2.25$ | $x = 3.00$ | $x = 3.75$ |
|---|---|---|---|---|---|---|
| Young's Modulus (Pa) $\mathbb{Y}_{Li,Si}$ | $5 \times 10^9$ | $6.3 \times 10^9$ | $8 \times 10^9$ | $9 \times 10^9$ | $10 \times 10^9$ | $12 \times 10^9$ |

**Figure 3.** The maximum volume increase at $\overline{n}_c = 6$ for each corresponding concentration $x$, which is equal to the ratio of the number of lithium ion to silicon atoms (Li/Si) for Li$_x$Si composite material. As the concentration $x$ decreases there is a corresponding decrease in the amount of volume change.

defined as

$$I_{\text{current}} = \frac{\overline{N}_c e^2 \boldsymbol{E}_{11}}{a m_{Li} \omega_{Li}} \quad \text{where} \quad \overline{N}_C = \overline{n}_c a^3 \qquad (6)$$

The current (**Figure 4**) is defined by the silicon lattice constant $a$, the electron charge constant $e$, the electric field $\boldsymbol{E}_{11}$ in the $\langle 110 \rangle$ direction, lithium ion mass $m_{Li}$, average negative charge differential $\overline{N}_C$, and the angular frequency $\omega_{Li}$ of the lithium ion.

With defining the current this leads to the derivation of the specific charge capacity $q_{scc}$ which is defined as

$$q_{scc} = \left(1 + \varepsilon_{11}\right)\frac{\overline{N}_\alpha}{\overline{N}_{Li}}\frac{e}{m_{Li}} \quad \text{where} \quad \overline{N}_\alpha = \alpha\overline{N}_C \tag{7}$$

$$\alpha = \frac{\omega_{ij}}{\omega_{Li}} \quad \text{where} \quad \omega_{ij} = \frac{eE_{11}}{\hbar\left(3\pi^2\overline{n}_c\right)^{\frac{1}{3}}}, \quad \omega_{Li} = \frac{\hbar\left(3\pi^2\overline{n}_{Li}\right)^{\frac{2}{3}}}{2m_{Li}} \tag{8}$$

where in addition to the parameters stated above for the current, $\overline{N}_{Li}$ is the number of lithium ions within the computational model [16]. The parameter alpha $\alpha$ is the ratio between the electron's angular frequencies $\omega_{ij}$ and the angular frequencies of the lithium ions. Most of the electrons are considered to be in the minimum conduction band of the quantum harmonic oscillator except for the electrons that are energetic enough to transition to a higher energy band [17]. However, this transition is temporary that last a fraction of a second as the electron emits a photon and returns to its previous state within the minimum conduction band. In this study the alpha parameter has been calculated at an average of $\overline{\alpha} = 10^3$ which is based on the range of $\alpha$ values between $\overline{n}_c = 2$ to $\overline{n}_c = 8$ and the average range of $x = 0.75$ to $x = 3.75$ of the $Li_xSi$ lithiated silicon material (**Figure 5**). This alpha of $10^3$ is when nc varies between 2 and 8 and when the average $x$ ratio of Li/Si is between 0.75 and 3.75 [18]. The average of alpha is also calculated for constant values of $n_c$ and Li$_x$Si in **Figure 4**. As an example if the average negative charge differential is constant at $\bar{n}_c = 5$ with a varying lithium ion diffusion rate between $x = 0.75$ and $x = 3.75$ than the corresponding average $\alpha$ is 1040. If the diffusion rate of Li$_x$Si is constant at $x = 1.50$ and average negative charge differential varies between $\bar{n}_c = 2$ and $\bar{n}_c = 8$ than $\bar{\alpha} = 1240$.

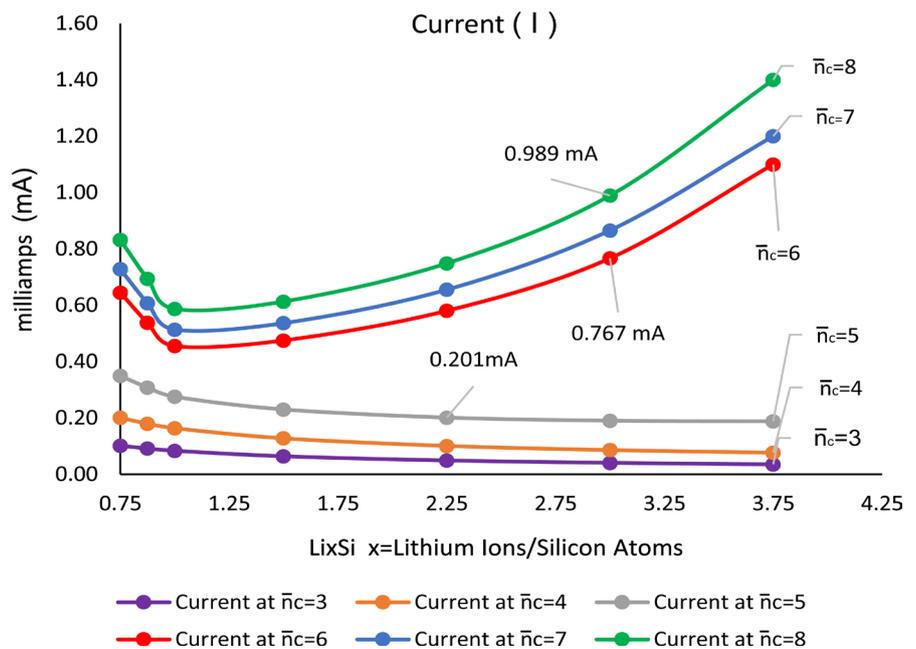

**Figure 4.** The current within the lithiated silicon nanowire for $\overline{n}_c < 6$ is approximately constant at any $x$ value. However at $\overline{n}_c \geq 6$ the electron flux experience an optical amplification through an increase of the electromagnetic field. The current increases nonlinearly as the concentration $x$ increases.

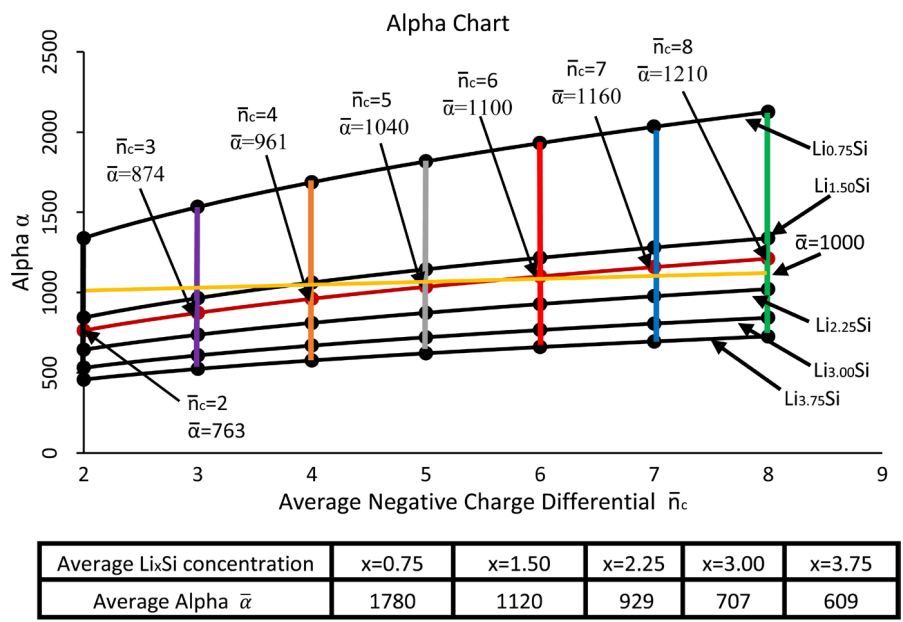

| Average Li$_x$Si concentration | x=0.75 | x=1.50 | x=2.25 | x=3.00 | x=3.75 |
|---|---|---|---|---|---|
| Average Alpha $\bar{\alpha}$ | 1780 | 1120 | 929 | 707 | 609 |

**Figure 5.** In this computational study $\alpha = 1000$ is used in Equations (7) and (8) and is defined as the average of $\alpha$ between the range of $\bar{n}_c = 2$ and $\bar{n}_c = 8$ and the average lithiated silicon Li$_x$Si range between $x = 0.75$ and $x = 3.75$. The average value of alpha $\bar{\alpha}$ is also given for the constant $\bar{n}_c = 2$ through $n_c = 8$ and for Li$_x$Si constant $x = 0.75$ through $x = 3.75$.

Alpha $\alpha$ can be interpreted as the velocity of the electron flux being greater than the velocity of the diffusion rate of the lithium ions by the parameter $\alpha$. This can be seen by the definition of the phase velocity of electron $v_e = \omega_{ij}/k$ versus phase velocity of lithium diffusion rate $v_{Li} = \omega_{Li}/k$ where $k$ is the wave number. The $q_{scc}$ is stated below in **Figure 5** and **Figure 6** versus the concentration $x = $ Li/Si and volume change $\Delta V_{11}$ respectively.

As diplayed in **Figure 6** and **Figure 7**, the experimental limit of 4200 mA-hr/g has been documented in the most recent literature and in this research study it has been chosen to be the maximum experimental capacity [9]. The computational findings in this research have shown that the specific charge capacity $q_{scc}$ can exceeded the experimental limit when $x < 1.00$. Since the specific charge capacity and current $I_{current}$ are average values and described by Poisson distribution, the of specific charge capacity standard deviations $\sigma_{scc}$ and the current standard deviation $\sigma_{current}$ are defined respectively as

$$\sigma_{scc} = \sqrt{q_{scc}} \; , \quad \sigma_{current} = \sqrt{I_{current}} \tag{9}$$

since the standard deviation value for the Poisson distribution is equal to square root of the mean.

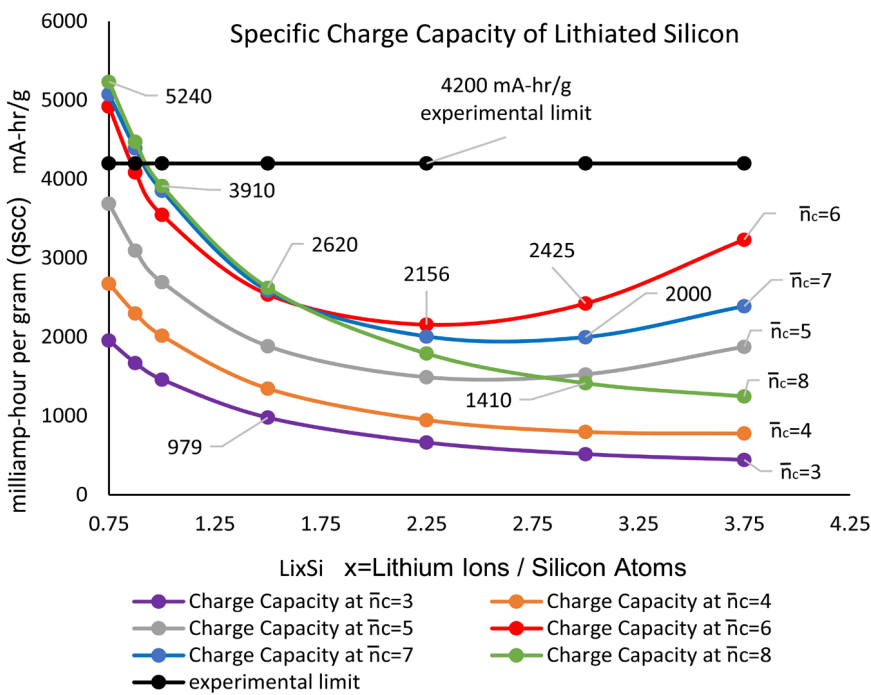

**Figure 6.** The highest specific charge capacity $q_{scc}$ is when $\bar{n}_c = 6$ is at $x \geq 2.25$, however at $x < 2.25$ the $q_{scc}$ steadily increases for all the $\bar{n}_c$ values. At $\bar{n}_c \leq 6$, the $q_{scc}$ at optical amplification reaches the experimental limit of 4200 milliamp-hour per grams.

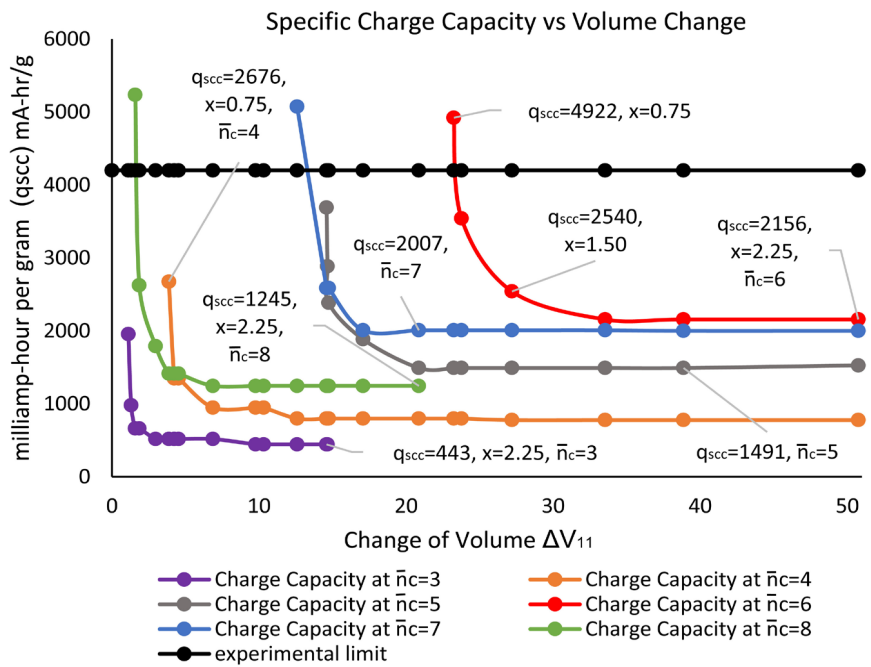

**Figure 7.** The specific charge capacity $q_{scc}$ is dependent on the change of volume. As it can be seen for each $\bar{n}_c$ value, the lower the volume increase correspond to a low concentration $x$. When $x < 1.00$, the $q_{scc}$ becomes exponentially large to the point of where the concentration is $x = 0.75$.

## 4. Conclusion

As was demonstrated, the specific charge capacity $q_{scc}$ of the lithiated silicon nanowire can be increased to experimental limits and possibly beyond if the concentration $x$ is reduced during lithiation process. With the reduction of the concentration $x$ also comes a decrease in the volume expansion. The increase in volume should be less than 50% increase of the original silicon nanowire volume. If it is possible to design a lithium ion battery where the lithium ion flux concentration rate can be regulated below $x = 2.25$ and at the same time work in concert with electron flux, crystalline silicon, c-Si, can be used as a great improvement in charge capacity for lithium ion batteries [19].

## Conflicts of Interest

The author declares no conflicts of interest regarding the publication of this paper.